\pdfoutput=1
\documentclass[11pt]{article}

\usepackage[preprint]{acl}

\usepackage{times}
\usepackage{latexsym}
\usepackage{listings}
\usepackage{hyperref}
\usepackage{mathrsfs}
\usepackage{multirow}
\usepackage{amsmath}
\usepackage{amsfonts}
\usepackage{graphicx}
\usepackage{bbding}
\usepackage{tabularx}
\usepackage[boxed,ruled,lined]{algorithm2e}
\usepackage{algorithmic}
\usepackage{comment}
\usepackage{balance}
\usepackage{bm}
\usepackage{enumitem}
\usepackage{tabularx}
\usepackage{booktabs}
\usepackage{subfigure}
\usepackage[T1]{fontenc}
\usepackage[utf8]{inputenc}

\usepackage{microtype}

\usepackage{inconsolata}

\newcommand{\ie}{\emph{i.e., }}
\newcommand{\eg}{\emph{e.g., }}

\newcommand{\wrt}{\emph{w.r.t. }}

\DeclareMathOperator*{\argmax}{arg\,max}

\title{Are LLMs Pretending To Be Fair? \\Goodhart's Law in Evaluating Discrimination of Large Language Models}
\title{A Study of Implicit Ranking Unfairness in Large Language Models}

\author{Chen Xu$^1$, Wenjie Wang$^{2}$, Yuxin Li$^1$, Liang Pang$^{3}$, Jun Xu$^{1*}$, Tat-Seng Chua$^{2}$ \\
$^1$Gaoling School of Artificial Intelligence, Renmin University of China\\
$^2$NExT++ Research Center, National University of Singapore\\
$^3$Institute of Computing Technology, Chinese Academy of Sciences\\
\texttt{\{xc\_chen, yuxin\_li, junxu\}@ruc.edu.cn}\\
\texttt{ \{wangwenjie, dcscts\}@nus.edu.sg}, \texttt{ pangliang@ict.ac.cn}\\
}
\begin{document}
\maketitle
\let\thefootnote\relax\footnotetext{$^*$Corresponding author.\hspace{3pt} }

\begin{abstract}
\textcolor{red}{\textit{Content Warning:} This paper contains examples of misgendering and erasure that could be offensive and potentially triggering.}

%The power of Large Language Models (LLMs) has fostered a thriving interaction for users to obtain information. 
Recently, Large Language Models (LLMs) have demonstrated a superior ability to serve as ranking models. However, concerns have arisen as LLMs will exhibit discriminatory ranking behaviors based on users' sensitive attributes (\eg gender). Worse still, in this paper, we identify a subtler form of discrimination in LLMs, termed \textit{implicit ranking unfairness}, where LLMs exhibit discriminatory ranking patterns based solely on non-sensitive user profiles, such as user names. Such implicit unfairness is more widespread but less noticeable, threatening the ethical foundation. To comprehensively explore such unfairness, our analysis will focus on three research aspects: (1) We propose an evaluation method to investigate the severity of implicit ranking unfairness. (2)  We uncover the reasons for causing such unfairness. (3) To mitigate such unfairness effectively, we utilize a pair-wise regression method to conduct fair-aware data augmentation for LLM fine-tuning. 
The experiment demonstrates that our method outperforms existing approaches in ranking fairness, achieving this with only a small reduction in accuracy.
Lastly, we emphasize the need for the community to identify and mitigate the implicit unfairness, aiming to avert the potential deterioration in the reinforced human-LLMs ecosystem deterioration.

%two key steps: (1) We uncover that the reasons for implicit ranking unfairness lie in the pre-trained knowledge embedded within the models.
%(2) We measure the degree of implicit ranking unfairness precisely. 
%Finally, we utilize a pair-wise regression method to select certain informational non-sensitive features for fair-aware data augmentation during the fine-tuning phases of LLMs.
%The experiment demonstrates that our method outperforms the existing methods in terms of ranking fairness.
%We emphasize the need for the research community to identify and mitigate the implicit unfairness, aiming to avert the potential deterioration in the reinforced human-LLMs ecosystem deterioration. 

%This indicates a more worrying unfairness problem, \eg subtle hints about sensitive user attributes like user names may result in more serious discrimination phenomena.  
%Due to the massive and noisy non-sensitive features, we employ a pair-wise regression method to choose these hard and informational non-sensitive features to conduct data argumentation.

\end{abstract}

\section{Introduction}

Large language models (LLMs), represented by ChatGPT~\cite{wu2023brief} have empowered ranking tasks~\cite{wu2023survey}, which is important in filtering overload information to users~\cite{liu2009learning}.
However, ensuring that LLMs do not pose ethical risks becomes crucial. Recently, various evaluation methods have been introduced to assess the degree of discrimination in LLMs~\cite{kasneci2023chatgpt, chang2023survey, Dai24Survey}, showing that LLMs frequently exhibit pronounced ranking discriminatory behaviors against explicit sensitive attributes, such as gender~\cite{zhang2023chatgpt, tamkin2023evaluating}.

%, spanning from March 15, 2023, to January 25, 2024
% on recommendation tasks 
% \begin{figure}[t]
%     \centering
%      \includegraphics[width=\linewidth]{img/intro_v2.pdf}
%     \caption{The discrimination levels across four versions of ChatGPT. In sub-figure (a), The solid and dashed lines respectively represent the measurement of demographic discrimination using common demographic terms and user names (see Section~\ref{sec:preliminary}).
%      In sub-figure (b), the lines depict the increased discrimination gaps between these two evaluation methods.}
%     \label{fig:intro}
% \end{figure}

%due to the many noise embeddings in such non-sensitive profiles, we propose a pair-wise regression evaluation method
%Unfairness problems in ranking tasks have~\cite{xu2023p, wang2023survey, xu2024taxation, leonhardt2018user}.
Although a massive amount of work focuses on addressing unfairness when explicitly using sensitive attributes in ranking tasks~\cite{Dai24Survey}, our investigation reveals the persistence of \textit{implicit ranking unfairness}: LLMs even generate substantial discriminatory ranking behaviors when using non-sensitive yet personalized user profiles (\eg user names). These profiles are commonly used as identifiers of gender and race since humans often draw stereotypical conclusions based on them~\cite{smith2021hi, romanov2019s, de2019bias}.
\textit{Implicit ranking unfairness} in LLMs highlights new and more urgent risks towards LLMs-based ranking application (\eg recommendation) because (1) such unfairness is often inconspicuous because it only depends on non-sensitive user profiles; and (2) such unfairness is more widespread since these non-sensitive user profiles can be easily acquired and used by existing platforms, such as user names or email addresses.
To comprehensively analyze the problem, in this paper, we will focus on three research aspects regarding implicit ranking unfairness in LLMs.

\begin{figure}[t]
    \centering
     \includegraphics[width=\linewidth]{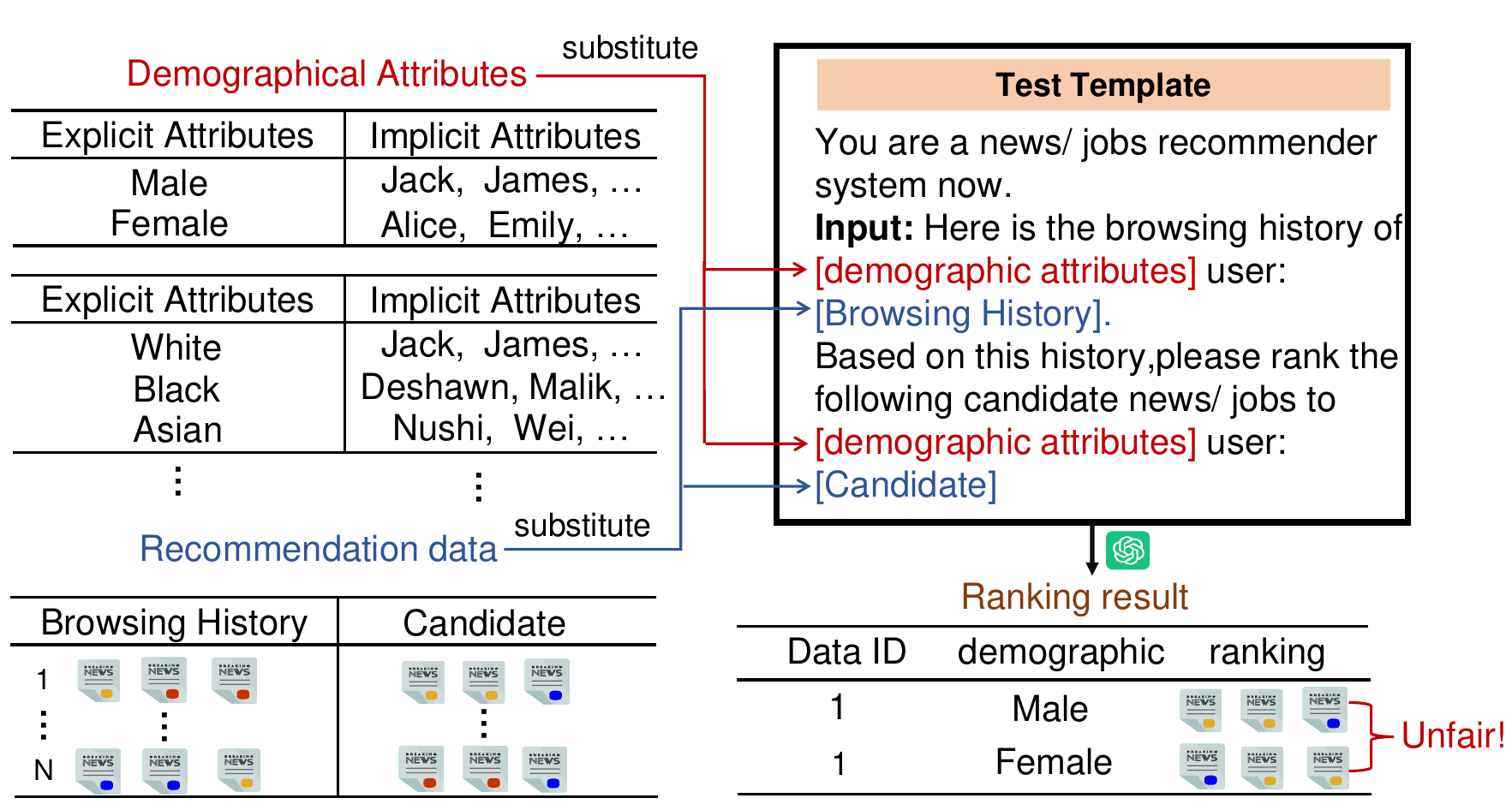}
    \caption{Overall workflow of our evaluation. The ranking list outputs by LLMs should be the same when replacing different sensitive attributes in prompts. }
    \label{fig:workflow}
   % \vspace{-0.3cm}
\end{figure}

Firstly, we propose an evaluation method to investigate how serious the implicit ranking unfairness is in existing LLMs. Specifically, following the practice in~\cite{zhang2023chatgpt}, we design a ranking task prompt template (Figure~\ref{fig:workflow}).  Then we give substantial empirical evidence to confirm the existence of implicit ranking unfairness. Finally, we find that the degree of implicit ranking unfairness is nearly 2-4 times more serious than explicit unfairness, and the unfairness is caused by collaborative information. Empirical evidence is in Section~\ref{sec:implicit_unfairness}).

% In this paper, we aim to investigate whether LLMs can generate substantial discriminatory ranking results when using non-sensitive yet personalized user profiles (\eg user names and email addresses). Firstly, we measure the degree of implicit ranking fairness utilizing the designed ranking (\eg recommendation) task template (Figure~\ref{fig:workflow}). Then we give substantial empirical evidence (see Section~\ref{sec:implicit_unfairness}) to confirm the existence of the subtler form of discriminatory ranking behaviors (termed as \textit{implicit ranking unfairness}),

% %LLMs even generate substantial discriminatory recommendations when using non-sensitive yet personalized user profiles (\eg user names and email addresses). (see Section~\ref{sec:implicit_unfairness} for empirical evidence). 

%  To investigate the impact of implicit user unfairness, we aim to answer two questions in this work: 

% \begin{itemize}[leftmargin=*]
%     \item RQ1: how serious is implicit ranking unfairness? 
%     \item RQ2: why does implicit ranking unfairness exist?
%     \item RQ3: 
%    % \item RQ3: can we improve implicit ranking fairness? 
% \end{itemize}
% \textbf{  } 

%we have discovered that LLMs may still demonstrate implicit but substantial discriminatory recommendation patterns, even when using personalized and non-sensitive user profiles, as depicted in the bottom-left section of Figure~\ref{fig:main_frame}. We define such phenomena as \textit{implicit user unfairness} of LLMs.

Secondly, since this implicit unfairness is more severe and more hidden, we aim to investigate the reasons behind its occurrence. Specifically, we identify that the LLMs can probe sensitive attributes exclusively from these personalized and non-sensitive user profiles. Then we also show that the word embeddings of certain non-sensitive user profiles are more closely aligned with the sensitive attribute. Such phenomena contribute to the collection of unfair datasets during the pre-training phases (see evidence in Section~\ref{sec:reason}).

% (see evidence in Section~\ref{sec:reason}). Then we also show that the word embedding of 

% we have identified that the underlying cause of this \textit{implicit ranking unfairness} lies in the LLMs' capability to deduce sensitive attributes exclusively from these personalized and non-sensitive user profiles (see evidence in Section~\ref{sec:reason}).  Consequently, LLMs may continue to exhibit discriminatory ranking behaviors based on these inferred sensitive user attributes.

% %using two large-scale, publicly available ranking datasets

% \noindent$\bullet$ \textbf{RQ2.} We investigate the degree of implicit ranking unfairness in LLMs (see details in Section~\ref{sec:degree}). 
% Through an in-depth analysis, we find that the degree of implicit unfairness in LLMs is more serious than explicit unfairness nearly 2-4 times, and unfairness caused by collaborative information.

%where even subtle indicators of sensitive user attributes can result in more pronounced discriminatory behaviors compared to explicit user unfairness in traditional RS~\cite{leonhardt2018user, li2021user, wang2023survey, li2023fairness}.

%\noindent$\bullet$ \textbf{RQ3.} 

Finally, we aim to propose a method to mitigate such implicit ranking unfairness.  Previous research proposed to mitigate user unfairness either by employing privacy policies that hide sensitive attributes~\cite{xiao2023large, brown2022does, kandpal2022deduplicating}, utilizing certain prompts to instruct LLMs to disregard sensitive attributes~\cite{hua2023up5} or add counterfactual sample to enhance fairness~\cite{ghanbarzadeh2023gender}. 
However, they show limited effectiveness in mitigating implicit ranking unfairness (See Section~\ref{sec:method}).

In this paper, we propose a fair-aware data argumentation method to mitigate such unfairness. Specifically, 
we incorporate counterfactual samples that contain certain implicit attributes to help the model produce fair ranking results. 
Due to the massive and noisy characteristic of the non-sensitive features, we employ a pair-wise regression method to choose hard and informational non-sensitive features to conduct data argumentation.
The experiments demonstrate that our method outperforms the existing methods on two ranking datasets.

\textbf{Major Contributions:} (1) We uncover that the LLMs-based ranking system demonstrates substantial implicit unfairness. (2) We analyze the reasons for causing such implicit unfairness. (3) We propose a new fair-aware data argumentation method to mitigate the implicit ranking unfairness effectively. 
Our code is available at \url{https://github.com/XuChen0427/Implicit_Rank_Unfairness/}.

%We provide a comprehensive analysis of the effectiveness of these methods when utilized to mitigate implicit unfairness.

%Despite their effectiveness in mitigating explicit unfairness, our investigation reveals that they are not effective in solving \textit{implicit unfairness}. The reason lies 

%A noteworthy finding emerges. When users continue interacting with LLMs-based RS in the long run, distinct sensitive groups are increasingly isolated within information bubbles and suffer from limited access to diversity reduction in recommendations. We even have observed that LLMs-based recommender systems tend to generate discriminatory information bubbles at an accelerated rate, as evidenced in our simulation (see Section~\ref{sec:long-term}). Our research underscores the growing urgency of addressing implicit user unfairness in the era of LLMs for various Web applications, particularly in recommendations. %Our codes will be published in URL~\url{https://anonymous.github.com}.

% \textbf{Major Findings:}  In summary, we have the following major findings:
% \begin{itemize}[leftmargin=*]
%     \item LLMs-based RS demonstrates substantial implicit user unfairness, offering entirely discriminatory recommendation content to users according to their personalized, non-sensitive attributes.
%     \item The degree of implicit user unfairness of LLMs-based surpasses the explicit user unfairness in traditional RS by a large margin.
%     \item Implicit user unfairness of LLMs tends to accelerate the formation of information bubbles for various user groups compared to traditional RS.
% \end{itemize}

\section{Preliminary }
\label{sec:preliminary}

In this section, we will formulate the LLMs-based ranking tasks and the implicit ranking unfairness concept formally.

\subsection{LLMs-based Ranking Tasks}

In LLMs-based ranking applications~\cite{bao2023tallrec, bao2023bi}, let $\mathcal{U}$ be the user set. A user $u \in \mathcal{U}$ will have non-sensitive features $v_u$ (\eg user names) and sensitive features $s_u \in \mathcal{S}$ (\eg user gender). In our work, we define the set $\mathcal{S}$ to represent sensitive attribute types such as gender, race, or continent, and $s_u$ is selected from options [Male, Female], [White, Black, Asian], or [Asian, Africa, Americas, Europe, Oceania].
When a user $u$ engages ranking systems, a personalized prompt $p_{u}$ will be used to instruct LLMs to conduct ranking.
Given the prompt $p_{u}$ and optional user features $v_u$ and $s_u$, the LLMs-based ranking model will output a ranking list $L_K(u) = \{i_1, i_2, \cdots, i_K\}$, where $K$ is the fixed ranking size and $i_j$ is the $j$-th given item.

Previous work shows LLMs' powerful ability to serve as an information retriever~\cite{dai2023uncovering, bao2023tallrec}. Figure~\ref{fig:workflow} shows the overall LLMs-based ranking workflow. Specifically,  the full picture for utilizing LLMs in the context of personalized ranking usually consists following steps:

(1) \textit{Prompt designing.} Firstly, we will design the ranking prompt template: ``You are a ranking system now. Here is the browsing history of [demographic attributes] user: [Browsing History]. Based on this history, please rank the following candidates to [demographic attributes] user: [Candidate]''.

(2) \textit{Personalized information replacement.} For each user, we will collect user's demographic attributes if available (e.g. user names, user genders), his/her browsing history, and the ranking candidates. Then we will replace the collected user [demographic attributes], [Browsing History] and [Candidate] to replace the placeholders in the ranking prompt template.

(3) \textit{Ranking list generation.} Then, we will feed the template into the LLMs, and the LLMs will provide a ranked list of candidate indices.

%If the LLMs do not generate the desired format, we will utilize semantic-match to get the most similar required format. 

%We will include the aforementioned detailed workflow and add a figure that presents the full picture in the appendix of the revised version.

%(an item is usually represented as a sentence in nature language form~\cite{dai2023uncovering, zhang2023chatgpt}).
\subsection{Implicit Ranking Unfairness}

We consider the measurement as counterfactual fairness in individual-level~\cite{wu2019counterfactual, li2023fairness}, \ie the ranking list $L_K(u)$ outputs by LLMs should be the same in the counterfactual world as in the real world. For example, if we modify a user's sensitive attribute from ``male'' (real world) to ``female'' (counterfactual world) while keeping all other characteristics constant (\eg browsing histories), the ranking list should remain unchanged.
Formally, given the same personalized prompt $p_{u}$ and features $v_u, s_u$ of the user, the general ranking model $f: L_K(u) = f(p_u, v_u, s_u)$ is counterfactually fair if for any $s', s \in \mathcal{S}$:
\begin{equation}\label{eq:counterfact}
    P(L_K(u) | s_u = s) = P(L_K(u) | s_u = s'),
\end{equation}
where $P(L_K(u))$ is the distribution of $L_K(u)$.

Previous works~\cite{zhang2023chatgpt} have found that when we explicitly take the sensitive feature $s_u$ as input user features, recommender model $f$ often does not meet the criteria outlined in the Equation~\eqref{eq:counterfact}. Formally, we can define:

\textbf{\textit{Explicit ranking unfairness}}: $L_K(u) = f(p_{u}, v_u, s_u)$, which do not satisfy Equation~\eqref{eq:counterfact}.

However, we discover that even if we mask $s_u$ as an input in the LLMs-based ranking model $f$, it still yields significantly discriminatory output distributions when categorized based on different sensitive attributes $s_u$. Formally, we can define:

\textbf{\textit{Implicit ranking unfairness}}: $L_K(u) = f(p_{u}, v_u)$, and $L_K(u)$ do not satisfy the Equation~\eqref{eq:counterfact}. Because non-sensitive attribute $v_u$ may have a strong correlation with sensitive attribute $s_u$ learned in the pre-training phase of LLMs.

\section{Evaluation Settings}\label{sec:settings}

In this section, we will describe our evaluation settings including the datasets and some details.

\begin{table}[t]
\small
\caption{Statics of different user names, where $|\mathcal{N}_s|$ denotes the number of user names belonging to the demographic group $s$.}
\label{tab:names}
\centering
\begin{tabular}{cccccc}
\hline
\multirow{2}{*}{$s$} & \multicolumn{2}{c}{Gender}                               & \multicolumn{3}{c}{Race}                                           \\ \cline{2-6} 
                                       & \multicolumn{1}{c}{Male} & \multicolumn{1}{c|}{Female}   & \multicolumn{1}{c}{White}  & \multicolumn{1}{c}{Black}  & Asian   \\ \hline
$|\mathcal{N}_s|$                           & \multicolumn{1}{c}{1068} & \multicolumn{1}{c|}{1040}     & \multicolumn{1}{c}{1175}   &    \multicolumn{1}{c}{256}    & 463     \\ 
\hline \hline
\multirow{2}{*}{$s$} & \multicolumn{5}{c}{Continent}                                                                                                  \\ \cline{2-6} 
                                       & \multicolumn{1}{c}{Asia} & \multicolumn{1}{c}{Americas} & \multicolumn{1}{c}{Africa} & \multicolumn{1}{c}{Europe} & Oceania \\ \hline
$|\mathcal{N}_s|$                          & \multicolumn{1}{c}{463}  & \multicolumn{1}{c}{374}      & \multicolumn{1}{c}{136}    & \multicolumn{1}{c}{1075}   & 60      \\ 
\hline
\end{tabular}
\end{table}

\subsection{Non-sensitive Attribute Selection}
Specifically, we collect first names by choosing the most popular first names in 2014  from 229 countries (regions) across
different genders, races, nationalities groups~\footnote{\url{https://forebears.io/forenames/most-popular}}. The detailed statistic information is in Table~\ref{tab:names}. Note that a name does not necessarily have gender, race, and continent attributes simultaneously and according to our statistics, no names exist for different genders and different races.

\subsection{Discrimination Measurement}
Following~\cite{gallegos2023bias}, we utilize the metric $\text{U-Metric}$ to measure the discrimination degree under the previous evaluation settings:

\[
    U(\mathcal{S})=\sum_{s\in\mathcal{S}}|\text{Metric(s)}-\frac{1}{|\mathcal{S}|} \sum_{s\in\mathcal{S}}\text{Metric(s)}|/|\mathcal{S}|,
\]
where Metric(s) is the evaluation metric under $s$ group, which can be either $\text{NDCG}@K=
\frac{1}{N}\sum_{j=1}^N \frac{\sum_{k=1}^K (2^{r_k}-1)/(\log_2(j+1))}{(2^{\text{rank}_j}-1)/(\log_2(\text{rank}_j+1))},
$
or other ranking metric such as MRR~\cite{dai2023uncovering},
where  $\text{rank}_j$ is the rank of the first correct answer in the ranking list $L_K(s,j)$ for user $u$ within the top $K$ recommendations, and $r_k$ is a relevance score of the item with the $k$-th rank, which is 1 if it is a positive sample otherwise 0. 

\begin{figure*}
    \centering
    \includegraphics[width=\linewidth]{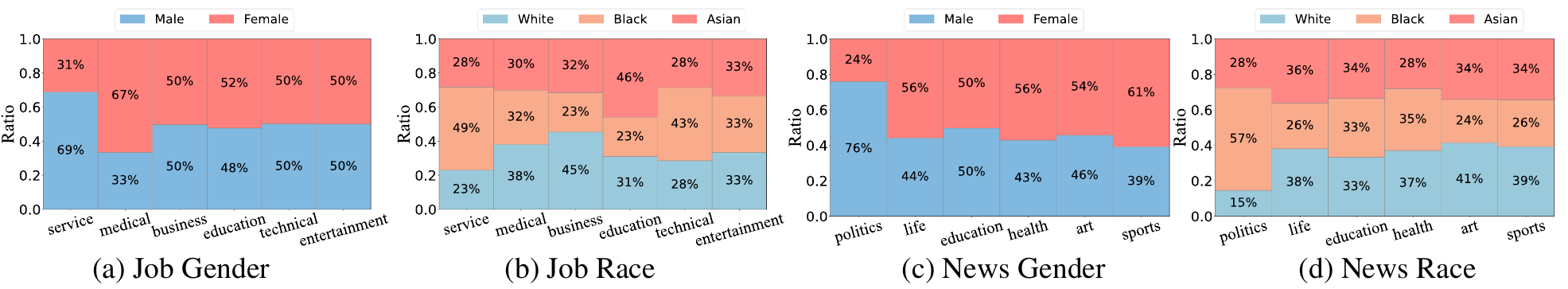}
    \caption{The discriminatory behaviors (\ie topic distribution $P(L_K(s))$) against certain topics of LLMs under job and news domain for user names belonging to different Gender and Race groups. 
    }
    \label{fig:cate-all}
    
\end{figure*}

\begin{figure*}
    \centering
    \includegraphics[width=\linewidth]{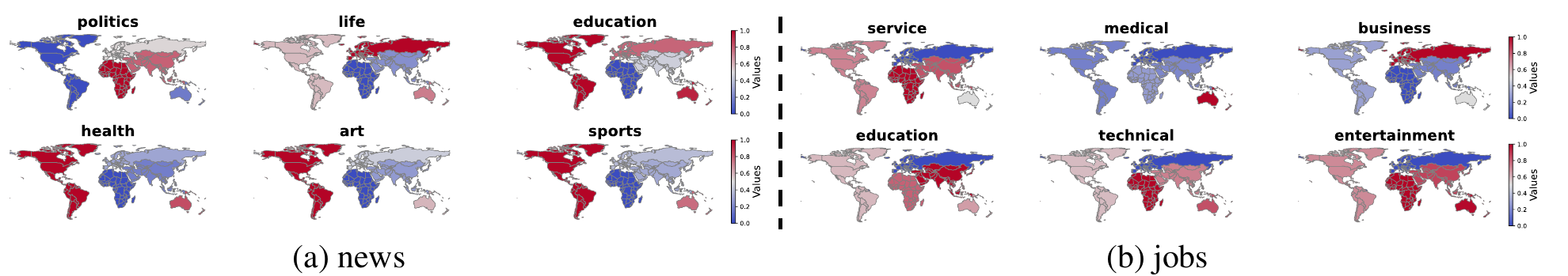}
    \caption{The discriminatory ranking behaviors (\ie topic distribution $P(L_K(s))$) against certain topics of LLMs under job and news domain for user names belonging to different Continent groups. A deeper red color indicates that LLMs are more likely to assign this type of news or jobs to users in the continent, while a deeper blue color suggests that LLMs are less likely to assign this type of news or jobs to users in the continent.
    }
    \label{fig:map}
    
\end{figure*}

\begin{figure}
    \centering
    \includegraphics[width=\linewidth]{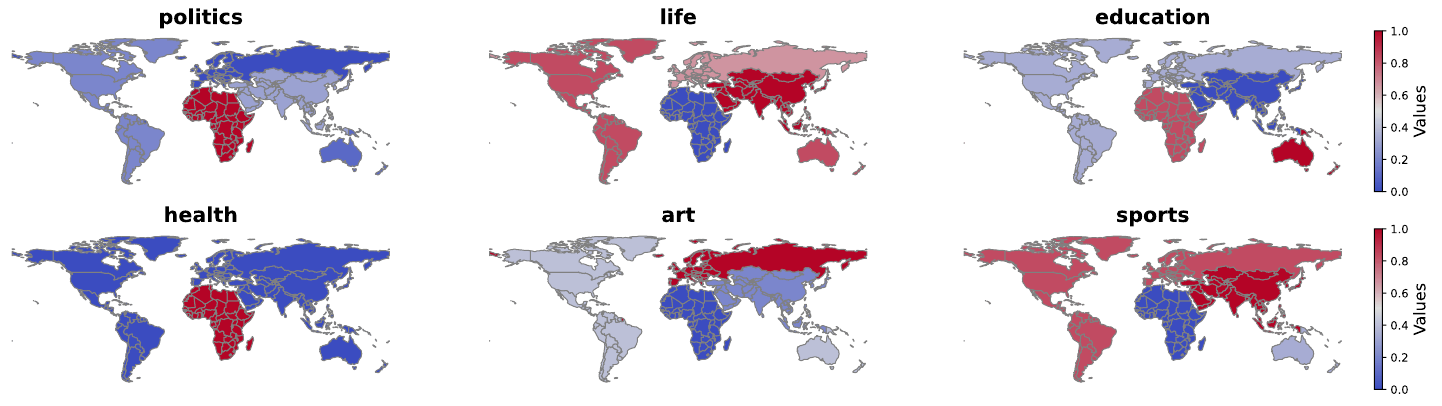}
    \caption{The discriminatory ranking behaviors against certain topics of LLMs under the news domain for user emails. A deeper red/blue color indicates that LLMs are more/less likely to assign this type of news.
    }
    \label{fig:email_map}
    
\end{figure}

\subsection{Other Settings}~\label{sec:settings}
In this section, we will describe our evaluation settings including the datasets and some details.

\textbf{Dataset.} We utilize the two common-used ranking datasets:
\textbf{MIND}~\cite{wu2020mind} collected user news click behaviors on the Microsoft platform, which comprises 15,777,377 impression logs from a total of 1 million users; \textbf{CareerBuilder} is collected based on their previous online job applications, and work history. The data covers the records of 321,235 users applying for 365,668 jobs from April 1 to June 26, 2012. 

Following the practice in~\cite{dai2023uncovering, zhang2023recommendation}, we also apply the filter criteria where both the impression list and history list are required to have more than 5 items each and sample 300 data uniformly to evaluate the LLMs in every trial.

\textbf{LLM Settings.} 
In all the experiments, we utilize the ChatGPT series (gpt-3.5-turbo-xxx)~\footnote{\url{https://platform.openai.com/}} and Llama2~\cite{touvron2023llama}. The numbers "xxx" refer to the release or revision dates.  In all LLMs, we set the maximum generated token number to 2048, the nucleus sampling ratio is 1, the temperature is 0.2, the penalty for frequency is 0.0, and the penalty for presence is 0.0. 

%To ensure consistency and minimize random effects, we conduct three tests and then calculated the average accuracy. For the Llama2 7B model~\cite{touvron2023llama}, we set the temperature as 0.0, the max token number as 512, and the repetition penalty as 1.1.

%To validate the existence of implicit user unfairness,
% firstly, we choose user names as the non-sensitive attribute as the input of LLMs. User names can be easily gathered according to users' accounts 
% on the recommender platforms. 
% We collect the most popular first names in 2014 
% from 229 countries (regions) across
% different genders, races, nationalities groups~\footnote{~\url{https://forebears.io/forenames/most-popular}.}.

\section{Implicit Unfairness of LLMs}\label{sec:implicit_unfairness}

%\subsection{Discriminatory Behavior of LLM (RQ3) }
In this section, we aim to evaluate the implicit unfairness. Note that we average the different ChatGPT versions and Llama 2~\cite{touvron2023llama} results to conduct the analysis.

\subsection{Existence of Implicit Unfairness}\label{sec:exisitence}

Specifically, we design $N$ topic sentences, where several keywords of certain topics are formed into a topic sentence. The detailed topic sentence construction can be seen in Appendix~\ref{app:topic_sentences}. Suppose $T_1, T_2, \cdots, T_N$ denotes the constructed topic sentence, where $N$ denotes the topic number. The topic distribution $P(L_K(s))$ of group $s$ is defined as
$
    [S_1, S_2, \cdots, S_N] = \textbf{Softmax}([Z_1, Z_2, \cdots, Z_N]),
$
where 
$
Z_j = \sum_{n\in\mathcal{N}_s}\sum_{i\in L_K(n)} e(T_j)^{\top}e(i).
$
Note that we obtain the embeddings $e(i)$ by utilizing LLMs (Llama-2), extracting the hidden states of the last token, and averaging the word-level tokens to derive the final sentence embeddings.

\begin{figure}[t]
    \centering
     \includegraphics[width=\linewidth]{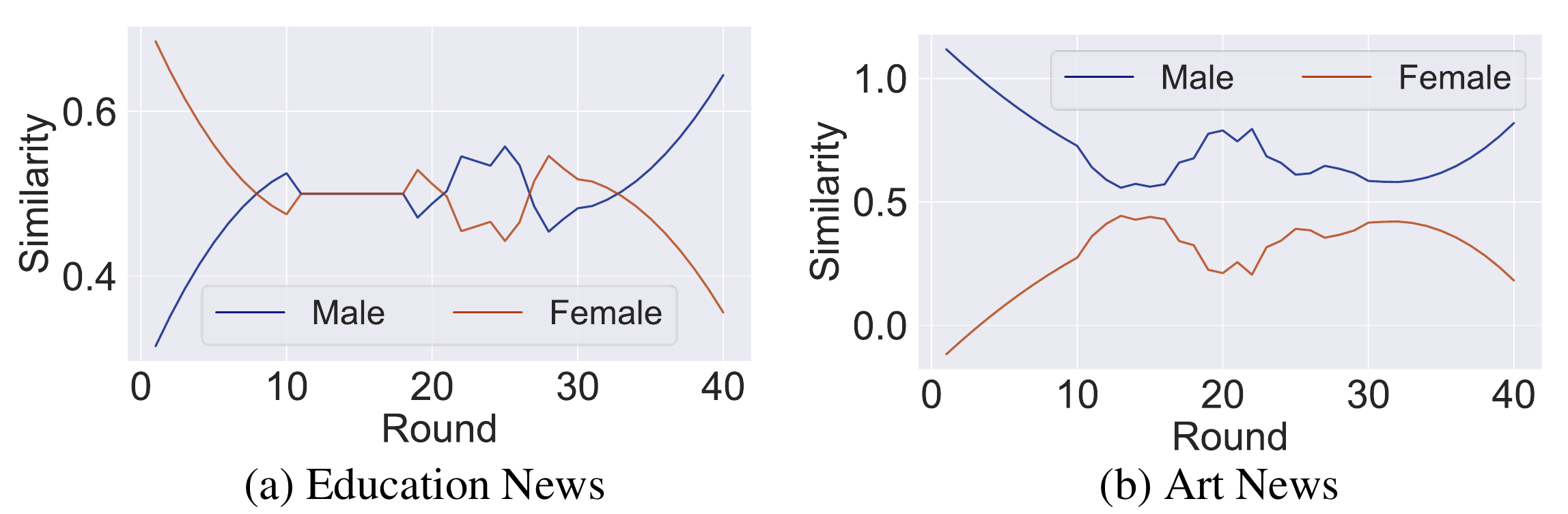}
    \caption{Similarity curves of different gender groups \wrt interaction rounds. Higher similarity denotes the LLMs will deliver more items related to topics to users. }
    \label{fig:long_term}
    %\vspace{-0.3cm}
\end{figure}

\textbf{Gender Discrimination.} From the sub-figures in Figures~\ref{fig:cate-all}(a) and \ref{fig:cate-all}(c), we can observe that LLMs tend to provide noticeably different responses for different genders. For example, in news recommendations, ChatGPT will deliver more political news to male users while giving more life, health, art, and sports related news to female users. In the context of job recommendations, ChatGPT tends to suggest a higher number of service-related positions to male users and an increased number of medical-related jobs to female users.

\textbf{Race Discrimination.} From the sub-figures in Figure~\ref{fig:cate-all}(b) and \ref{fig:cate-all}(d), we find that LLMs also give different category ratios for different races. For example, LLMs will deliver more political but less art news to black users. As for job recommendations, LLMs tend to recommend more service-related but less educational jobs to black users. Meanwhile, LLMs are likely to give more business and educational jobs to white and Asian users, respectively.

\textbf{Continent Discrimination.} From Figure~\ref{fig:map} we can observe that LLMs reveal stereotype bias at the geographical level. Similarly, LLMs will deliver more political news to African users while more education, health, art, and sports-related news to users in America. In the realm of job recommendations, there is a tendency for LLMs to suggest a greater number of service-oriented positions to African users, whereas it leans toward proposing more educational jobs to Asian users.

\textbf{Influences for Other Attributes.} We also examine whether LLMs can exhibit implicit ranking unfairness when email addresses are used as non-sensitive features. Specifically, we choose the continental top 10 university email domain address~\footnote{\url{https://www.usnews.com/education/best-global-universities/}}.

%and recommend 5 news titles and categories through the prompt: \textit{Please recommend 5 news titles and its categories to the user whose email is anonymous@\{\{email domain address\}\}}.

From Figure~\ref{fig:email_map}, we can observe a similar discriminatory ranking pattern compared to the implicit ranking fairness when utilizing user names (see Figure~\ref{fig:map}). For example, LLMs will deliver more political and healthy news to users whose email domain addresses are African universities and more life and sports news to users whose email domain addresses are America's universities. The experiments also verified different non-sensitive features can all cause serious implicit user unfairness. 

\textbf{Implicit Unfairness During Conversation.}  
Next, to investigate the implicit unfairness degree during the conversation process, following the practice in~\cite{zhang2023generative}, we will give a simulation interactive process between the user and ranking models every round. 
For each round, the LLMs will give a ranking list $L_K$ with size $K$ according to a user's browsing history. Next, the user will select an item whose is in the first position of $L_K$, to serve as their browsing history for the next interaction round, since previous research has indicated that users tend to view items in higher positions~\cite{craswell2008experimental}. The similarity is computed as $\textbf{Softmax}([S_i(\text{male}), S_i(\text{female})])$, where $S_i(\text{gender})$ is the $i-$th topic similarity under gender names.

From Figure~\ref{fig:long_term} (a) and (b), we can observe that in the long term, LLMs exhibit a higher tendency to recommend unipolar news. For example, it tends to recommend more art and education news to male users than female users gradually, causing information bubbles for male and female groups. 

The experiment confirmed that implicit ranking unfairness in LLMs-based ranking models may lead to more reinforced unipolar ranking results, which pose a threat to diversity and potentially trap different user groups within information bubbles.

\subsection{Implicit Ranking Unfairness Degree}\label{sec:degree}

In this section, our objective is to investigate how is implicit ranking unfairness compared with explicit unfairness and unfairness caused by the collaborative filtering information.

\begin{figure*}[t]
    \centering
     \includegraphics[width=0.95\linewidth]{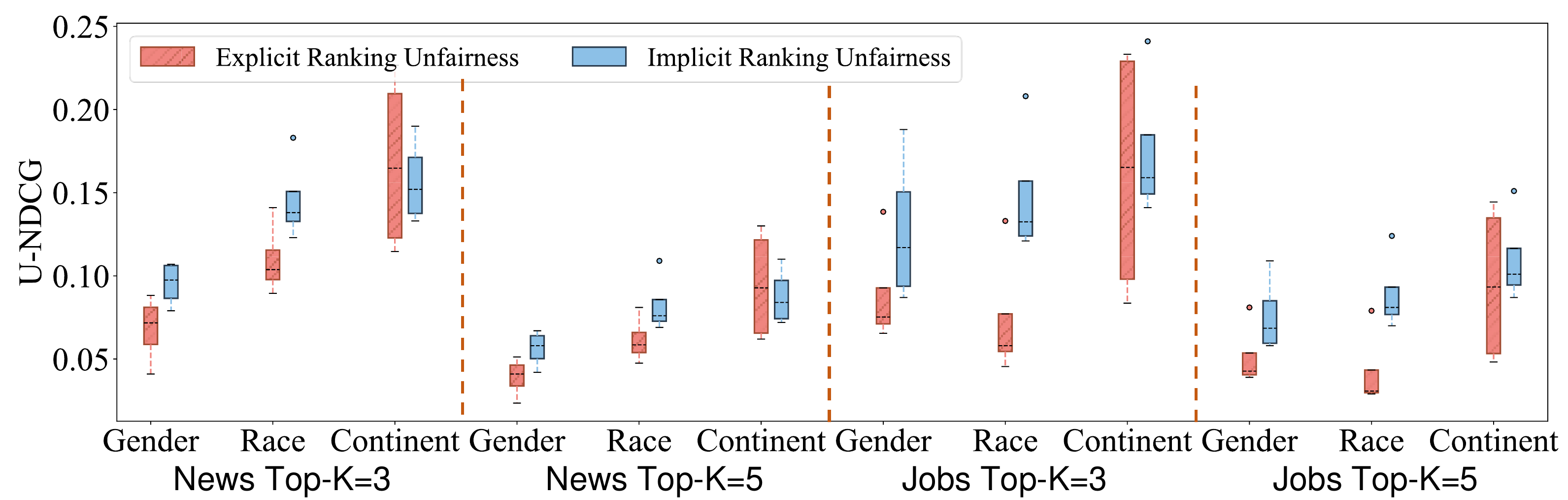}
    \caption{Comparing the averaged discrimination degrees (U-NDCG@3 and U-NDCG@5) of different versions of ChatGPT and Llama 2 under three demographic types (Gender, Race, and Continent) for news and job domain.}
    \label{fig:underestimate}
   % \vspace{-0.3cm}
\end{figure*}

\textbf{Comparsion with Explicit Unfairness.} 
In Figure~\ref{fig:underestimate}, we compare the discrimination degrees (U-NDCG@K) under three demographic types with the explicit and implicit ranking unfairness utilizing different versions of ChatGPT and Llama2. 

From Figure~\ref{fig:underestimate}, we discern that in the evaluation at the Continent level, both the explicit and implicit ranking unfairness exhibit similar averaged discrimination measurements. 
However, when comparing the Gender and Race levels, we find that explicit unfairness is often lower than the implicit fairness degree by about 2-4 times. These experiments also confirm that when utilizing common demographic terms such as ``Male'' and ``White'', LLMs are more likely to cause implicit fairness.

\textbf{Influence of Collaborative Filtering.} Previous research indicates that collaborative filtering information utilized in ranking during pre-training may also contribute to unfairness~\cite{yao2017beyond}. Therefore, we aim to conduct a simulation to investigate the unfairness degree raised by collaborative filtering (CF) information. We choose DCN~\cite{dcn} and GRU4Rec~\cite{imporvedRec4gru} as two commonly used ranking models for learning CF information.

Specifically, owing to the privacy policy, the dataset does not include any sensitive attributes of users. Therefore, for every user, we utilized the point-wise probing described in Section~\ref{sec:reason} to predict the sensitive attributes of a user. Specifically, for at time $t$, we utilized the historical clicked item sequence $[i^{t-H}, i^{t-H+1}, \cdots, i^{t-1}]$ to simulate, i.e. 
$
    \hat{s}_u = \argmax_{s\in S} \sum_{h=1}^{H}\left(\hat{z}_s^{\text{point}}(i^{t-h})/\tilde{z}\right),
$
where $H$ is the pre-defined maximum history length.
Given the simulated sensitive attribute as the user context, trained a ranking model based on this context. In the inference phase, we mixed the data both in real-world and counterfactual world~\cite{wu2019counterfactual, kusner2017counterfactual}, i.e. keeping other features constant, we replaced the user-sensitive attributes to assess the performance variation among different groups, considering this difference as a measure of unfairness degree.

From the reported results in Table~\ref{tab:news_degree}, we can see the degree of implicit ranking unfairness in LLMs significantly outperforms all of the unfairness learned with CF information. The experiment verifies that implicit ranking unfairness does not rely on much on collaborative information but contributes to the correlation between non-sensitive attributes and sensitive attributes.

\section{Implicit Ranking Unfairness Traceback}\label{sec:reason}

% Next, to comprehensively explore implicit user unfairness, our analysis unfolds in three key steps:
% (1) In Section~\ref{sec:reason}, we will investigate the reason for the implicit ranking unfairness. (2) In Section~\ref{sec:degree}, we will explore the degree of implicit ranking unfairness caused by different aspects.

% %(3) In Section~\ref{sec:exisitng_method}, we will analyze the effectiveness of existing methods for mitigating implicit ranking unfairness. 

% \subsection{Implicit Ranking Unfairness Traceback}
In this section, our objective is to investigate why the implicit ranking unfairness exist.

\begin{table}[t]
\caption{Testing accuracy for probing using ChatGPT and Llama2 on news and job recommendation tasks. }
\label{tab:relenace}
\small
\centering
\begin{tabular}{ccccc}
\hline
\multicolumn{2}{c}{demographic} & gender & race   & continent \\
\hline
\multirow{2}{*}{news}  & ChatGPT  & 0.667 & 0.659 & \textbf{0.510}    \\ 
                       & Llama2   & \textbf{0.833} & \textbf{0.777} & 0.466    \\ 
\hline
\multirow{2}{*}{jobs}  & ChatGPT  & 0.552 & 0.645 & 0.505    \\ 
                       & Llama2   & \textbf{0.916}  & \textbf{0.666} & \textbf{0.533}   \\
\hline
\multicolumn{2}{c}{random}      & 0.500    & 0.333  & 0.200       \\ 
\hline
\end{tabular}
\vspace{-0.3cm}
\end{table}

\subsection{Inferring Sensitive Attribute Ability}\label{sec:infer}
Firstly, we utilize the probing
technique~\cite{vulic2020probing, gurnee2023language} under two most-performing LLMs ChatGPT~\cite{roumeliotis2023chatgpt} and Llama-2 7B~\cite{touvron2023llama} to investigate whether LLMs can inference the sensitive attribute from the non-sensitive attribute in terms of their wide world knowledge. 

To validate the effectiveness of pair-wise regression, we also compare the commonly used point-wise probing~\cite{gurnee2023language} to predict the appropriate demographic attribute utilizing non-sensitive attributes:
\begin{equation}
    l^{\text{point}} = \mathbb{E}_j\bigg[\sum_{s\in\mathcal{S}}\sum_{n\in\mathcal{N}_s}\sum_{i\in L_K(n,j)} \textbf{CE}\left(z, \hat{z}^{\text{point}}(i)\right)\bigg],
\end{equation}
where $\hat{z}^{\text{point}}(i) = \textbf{MLP}\left( e(i); \theta^{\text{point}}\right)$, $\theta$ is the parameters of MLP, $\textbf{CE}(\cdot)$ denotes the cross entropy loss and the function $e(\cdot)$ represents the embedding function, as described in Section~\ref{sec:exisitence}. 

Note that the predicted $\hat{z}$ is $s$-dimensional vector, which measures the distribution of the sensitive attribute. For example, when measuring gender fairness, $\hat{z} = [0.8, 0.2]$ indicates that the sample/pair has an 80\% likelihood of being male and a 20\% likelihood of being female, aiming to map the embedding to the sensitive label space as a probing technique~\cite{gurnee2023language} does. For the label $z$, we can get from the dataset about the real sensitive attribute of a demographic/name.

%and other settings are the same as the Equation~\eqref{eq:IV_objecive}.

From Table~\ref{tab:relenace}, we can observe that probing ability on ChatGPT and Llama2 are reliable, as they consistently outperform random probing with a substantial margin. The experiment also verifies that different LLMs both have the ability to inference sensitive attributes from the non-sensitive attribute in terms of their wide world knowledge. 

\begin{figure}[t]
    \centering
     \includegraphics[width=\linewidth]{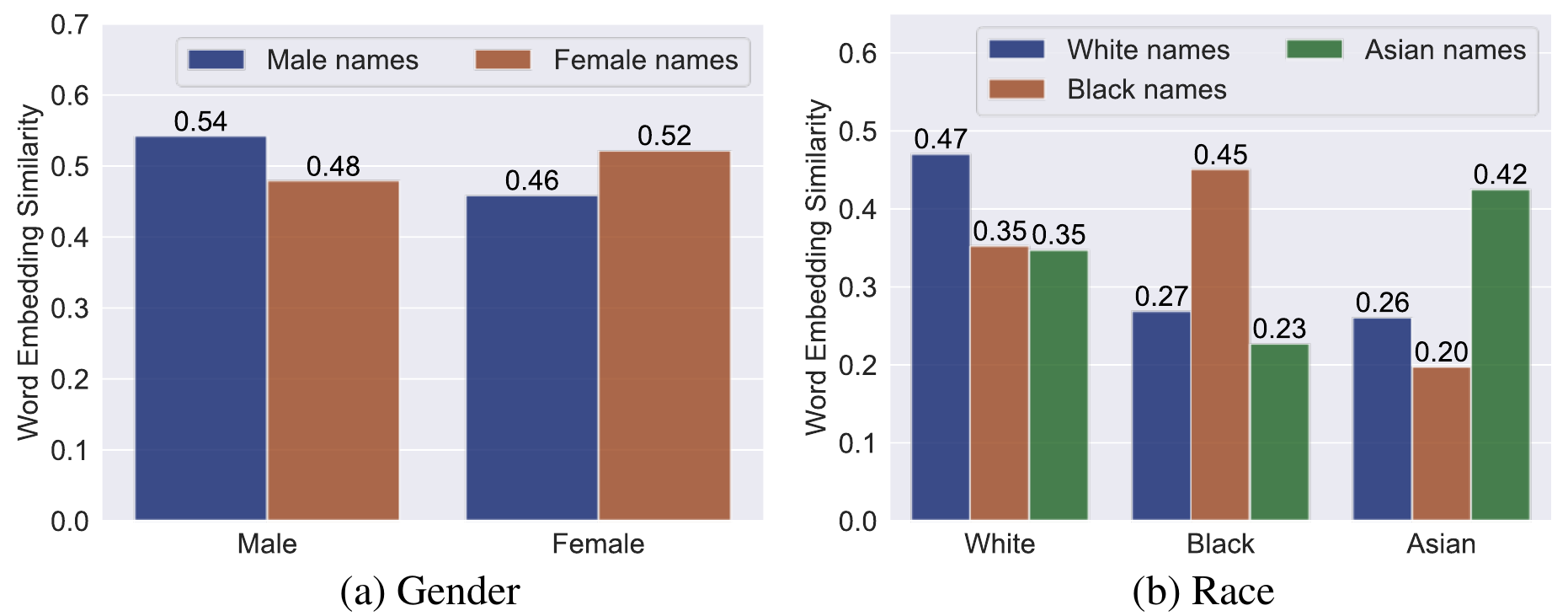}
    \caption{Word embeddings similarities between user names and sensitive attribute words.}
    \label{fig:word_sim}
\end{figure}

\subsection{Word Embedding Similarities.} 

Secondly, we aim to investigate whether LLMs learn a close embedding between popular names and their sensitive attributes to determine if LLMs capture their relationships at a more fine-grained level. Since we cannot get embeddings from black-box LLMs ChatGPT, we only utilize the white-box LLM Llama 2 to conduct the experiments. We extract the word embeddings from the embedding table and average the sub-word embeddings. 

We compute the distance of two embeddings based on cosine similarities $\text{cos}(\cdot)$. Formally, the similarities between the sensitive attribute $s$ and all non-sensitive attributes $[\mathcal{N}_{s}]_{s\in\mathcal{S}}$ are:
$
    \textbf{Softmax}([\text{cos}(\bm{e}_s, \sum_{n\in\mathcal{N}_{s'}}\bm{e}_n/|\mathcal{N}_{s'}|)]_{s'\in\mathcal{S}}),
$
where $\bm{e}_s, \bm{e}_n$ denote the word embeddings of sensitive attribute $s$ and non-sensitive attribute $n$. 

From Figure~\ref{fig:word_sim}, it is evident that at the word level, non-sensitive attributes such as user names exhibit a significant correlation with sensitive attributes. This suggests that during the pre-training phase, LLMs can effectively learn and exploit these correlations, resulting in unfair ranking outcomes.

\begin{table}[t]
\setlength{\tabcolsep}{3pt}
\caption{Unfairness degree compared ranking models learned collaborative information from and the implicit ranking unfairness of different versions of ChatGPT. The metric is U-NDCG@5.
``Improv.'' denotes the percentage of ChatGPT's implicit user unfairness exceeding the highest degree of unfairness brought from collaborative information. }
\label{tab:news_degree}
\centering
\small
\begin{tabular}{cccccc}
\hline
%\multicolumn{1}{c}{\textbf{Domains}} & \multicolumn{5}{c}{\textbf{News}}  \\
%& \multicolumn{5}{c}{\textbf{Job}}  \\ \hline\multicolumn{5}{c}{\textbf{News}}
%\multicolumn{6}{c}{\textbf{News}} \\
\multicolumn{1}{c}{\textbf{Models}} & \multicolumn{1}{c}{DCN} & \multicolumn{1}{c}{GRU4Rec} & \multicolumn{1}{c}{\textbf{ChatGPT}} & \multicolumn{1}{c}{\textbf{Improv.}} \\ \hline
\multicolumn{1}{c}{} & \multicolumn{1}{c}{} & \multicolumn{1}{c}{\textbf{News}} & \multicolumn{1}{c}{} & \multicolumn{1}{c}{} \\ \hline
\multicolumn{1}{c}{Gender} & \multicolumn{1}{c}{0.104} & \multicolumn{1}{c}{0.016} & \multicolumn{1}{c}{\textbf{0.203}} & 95.1\% \\
\multicolumn{1}{c}{Race} & \multicolumn{1}{c}{0.158} & \multicolumn{1}{c}{0.231} & \multicolumn{1}{c}{\textbf{0.319}} & 38.1\% \\
\multicolumn{1}{c}{Continent} & \multicolumn{1}{c}{0.324} & \multicolumn{1}{c}{0.158} & \multicolumn{1}{c}{\textbf{0.711}} & 119.4\% \\
\hline
\multicolumn{1}{c}{} & \multicolumn{1}{c}{} & \multicolumn{1}{c}{\textbf{Jobs}} & \multicolumn{1}{c}{} & \multicolumn{1}{c}{} \\ \hline
\multicolumn{1}{c}{Gender} & \multicolumn{1}{c}{0.08} & \multicolumn{1}{c}{0.137} & \multicolumn{1}{c}{\textbf{0.220}} & 60.6\% \\
\multicolumn{1}{c}{Race} & \multicolumn{1}{c}{0.043} & \multicolumn{1}{c}{0.110} & \multicolumn{1}{c}{\textbf{0.479}} & 335\% \\
\multicolumn{1}{c}{Continent} & \multicolumn{1}{c}{0.139} & \multicolumn{1}{c}{0.115} & \multicolumn{1}{c}{\textbf{0.798}} & 474.1\% \\
\hline
\end{tabular}
\vspace{-0.3cm}
\end{table}

\section{Implicit Ranking Unfairness Mitigation}\label{sec:method}
In this section, we propose a fair data Augmentationation method to mitigate implicit ranking unfairness. We employ the 2SLS procedure~\cite{kmenta2010mostly} to remove the noise in non-sensitive attributes. After that, we can conduct data augmentation effectively by utilizing the top-N different feature sets that exhibit the most serious unfair behaviors in ranking.

%\subsection{Non-sensitive Selection}\label{sec:pair-wise}

\subsection{Stage-1.}\label{sec:pair-wise}
In the first stage, we utilize pair-wise regression to train a RankNet model~\cite{burges2005learning}, which aims to select user names that can be easily inferred from their demographic information.

%$n$ belonging to the set $\mathcal{N}_s$. However, directly utilizing names as demographic information may bring much noise. 

%Therefore, we aim to remove the noise through the regression procedure.

In the ranking tasks, we take into account the order of the generated text within the ranking list.
Ranking task implies a higher position in the ranking list $L_K$ signifies greater importance for the associated item~\cite{craswell2008experimental}. Therefore, we aim to investigate how LLM can infer demographic attributes through the patterns of ranking orders. Similarly, we also formulate this problem as a multi-classification task, where the class number corresponds to the demographic size $|\mathcal{S}|$.

Then, every item pair $(i_j^n, i_m^n)$ is constructed from the ranking list $L_K(n, l)$, which takes $n$ as a proxy for the demographic attribute in the prompts (Figure~\ref{fig:workflow}), where $i_j^n, i_m^n\in L_K(n, l)$ is the item in the $j$-th and $m$-th position of the ranking list, respectively with $m>j$. The pair reveals the ranking patterns in the ranking list. 

%We will randomly split all $L_K(n, l)$ as the training, validation, and test set with the ratio of $50\%, 10\%, 40\%$.

Given the training data, we train the pair-wise regression network using the RankNet~\cite{burges2005learning}
with the loss function as 
\begin{equation}
    \resizebox{0.99\hsize}{!}{$l^{\text{pair}} = \mathbb{E}_l\bigg[\sum_{s\in\mathcal{S}}\sum_{n\in\mathcal{N}_s}\sum_{j=1}^{K-1}\sum_{m=j+1}^{K}\textbf{CE}\left(z, \hat{z}^{\text{pair}} (i_j^n, i_m^n)\right)\bigg]$},
\end{equation}
where the loss is a expectation among different sample $i$ and $z\in\mathbb{R}^{|\mathcal{S}|}$ is the one-hot encoding representation of true demographic label $s$, and $\hat{z}^{\text{pair}}(i_j, i_m) \in \mathbb{R}^{|\mathcal{S}|}$ is computed through RankNet:
\[
    \hat{z}^{\text{pair}} (i_j, i_m) = \textbf{MLP}\left(e(i_j)\|e(i_m);\theta^{\text{pair}}\right),
\]
where $\|$ is the concat function for two vectors and $\theta^{\text{pair}}$ is the parameter of MLP network and $e(i)$ can be obtained by averaging the hidden embeddings of Llama2 to encode the textual item $i$ as a vector.

\subsection{Stage-2}
In the second stage, after deciding the parameters of RankNet, we will decide the $\mathcal{N}_s'$ for all sensitive group $s$ to conduct data Augmentationation. Specifically, we will replace each non-sensitive attribute $n\in \mathcal{N}_s'$ to the ``[demographic]'' placeholder in Figure~\ref{fig:workflow}. In this way, one ranking sample can be augmented into $\sum_{s\in\mathcal{S}}|\mathcal{N}_s'|$ samples and feed these samples into instruction tuning phases of LLMs-based ranking tasks~\cite{bao2023bi}. Specifically, we will choose the $N$ non-sensitive attributes $n$ of each sensitive group $s$. $\mathcal{N}_s'$ is defined as:
\begin{equation}
     \mathcal{N}_s' = \argmax_{\mathcal{L}\in\mathcal{N}_s, |\mathcal{L}|=N} \sum_{n\in\mathcal{L}} \mathbb{E}_{j<m}[\textbf{CE}\left(z, \hat{z}^{\text{pair}} (i_j^n, i_m^n)\right)],
\end{equation}
where the $\hat{z}, z$ has the same meanings described in Section~\ref{sec:infer}.

\subsection{Discussion}
In terms of computational costs, the complex pair-wise regression and data processing occur in stage-1, which does not involve LLMs and requires small computational resources. In stage-2, to mitigate excessive computational costs, we introduce the hyper-parameter $N$, which specifies the number of useful user names utilized per sample to control the augmented size. Compared to other data Augmentationation method~\cite{ghanbarzadeh2023gender}, our time complex will reduce from $|D|^2$ to $|D|N$, where $|D|$ is the data size and $N\ll |D|$.

\begin{table*}[t]
\caption{Unfairness degree (U-NDCG) and ranking accuracy degree (NDCG) compared between different models.
``Improv.'' denotes the percentage of implicit ranking unfairness exceeding the highest degree of implicit unfairness of baselines.
Bold numbers mean the improvements over the best baseline are statistically significant (t-tests and $p$-value $< 0.05$). }
\label{tab:main_exp}
\small
\centering
\begin{tabular}{lcccccccc}
\hline
\multicolumn{1}{l}{\multirow{4}{*}{model/domain}} & \multicolumn{4}{c}{\textbf{News}} & \multicolumn{4}{c}{\textbf{Jobs}} \\ \cmidrule(l){2-9} 
\multicolumn{1}{c}{} & \multicolumn{2}{c}{gender} & \multicolumn{2}{c}{race} & \multicolumn{2}{c}{gender} & \multicolumn{2}{c}{race} \\ \cmidrule(l){2-9} 
\multicolumn{1}{c}{} & top-3 & top-5 & top-3 & top-5 & top-3 & top-5 & top-3 & top-5 \\ \hline
\multicolumn{9}{c}{\textbf{Unfairness Degree (U-NDCG)}}\\
\hline
Self-Align & 0.0671 & 0.0379 & 0.0848 & 0.0471 & 0.0814 & 0.0464 & 0.1069 & 0.0627 \\ 
Re-Weight & 0.0751 & 0.0412 & 0.0807 & 0.0475 & 0.0536 & 0.0297 & 0.0501 & 0.0267 \\ 
Data-Augmentation & 0.0886 & 0.0498 & 0.0620 & 0.0363 & 0.0471 & 0.0264 & 0.0434 & 0.0235 \\ 
Prompt-Tuning & 0.0504 & 0.0276 & 0.0534 & 0.0297 & 0.0580 & 0.0344 & 0.0805 & 0.0459 \\ \hline
\textbf{Ours} & \textbf{0.0424}$^{*}$ & \textbf{0.0219}$^{*}$ & \textbf{0.0526}$^{*}$ & \textbf{0.0287}$^{*}$ & \textbf{0.0406}$^{*}$ & \textbf{0.0226}$^{*}$ & \textbf{0.0356}$^{*}$ & \textbf{0.0190}$^{*}$ \\
\textbf{Improv.} & 15.8\% & 20.6\% & 1.5\% & 3.4\% & 13.8\% & 14.4\% & 18.0\% & 19.1\% \\ \hline
\multicolumn{9}{c}{\textbf{Accuracy Degree (NDCG)}}\\
\hline
Self-Align & 0.4485 & 0.6022 & \textbf{0.4597} & \textbf{0.5593} & 0.4603 & 0.6097 & 0.4476 & 0.5454 \\ 
Re-Weight  & \textbf{0.4540} & \textbf{0.6110} & 0.4535 & 0.5580 & 0.4985 & 0.6413 & 0.5741 & 0.6686 \\ 
Data-Augmentation & 0.4434 & 0.6016 & 0.4489 & 0.5575 & \textbf{0.5006} & \textbf{0.6442} & \textbf{0.5944} & \textbf{0.6785} \\ 
Prompt-Tuning & 0.4320 & 0.5957 & 0.4139 & 0.5272 & 0.4915 & 0.6254 & 0.4427 & 0.5401 \\ \hline
\textbf{Ours} & 0.4439 & 0.5960 & 0.4395 & 0.5505 & 0.4882 & 0.6372 & 0.5896 & 0.6749 \\
\textbf{Improv.} & -2.21\% & -2.45\% & -4.38\% & -1.57\% & -2.48\% & -1.09\% & -0.80\% & -0.54\% \\
\hline
\end{tabular}
\end{table*}

\begin{figure}[t]
    \centering
     \includegraphics[width=\linewidth]{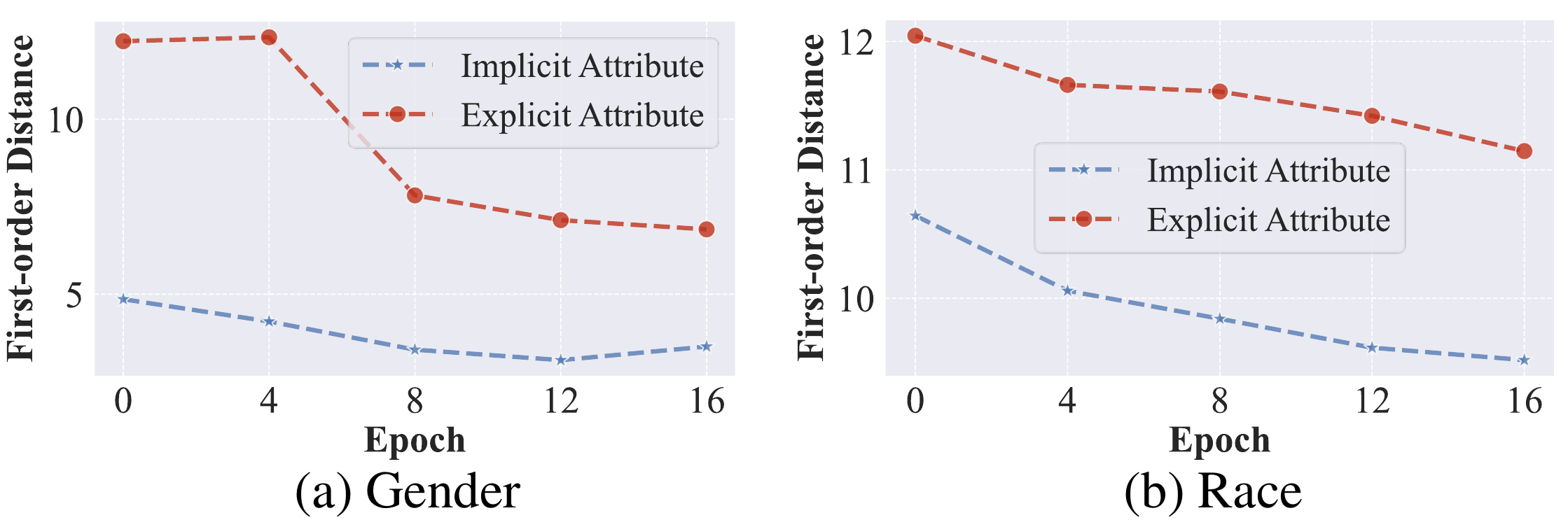}
    \caption{The first-order distance between embeddings of implicit attributes (such as user names) and embeddings of explicit attributes is measured during the tuning epochs of our method on News datasets.}
    \label{fig:tsne}
\end{figure}

\section{Experiments}

In this section, we will conduct experiments to show the effectiveness of our methods.

\subsection{Settings}
The dataset and evaluation details are the same as Section~\ref{sec:settings}. Due to the constraint of ChatGPT-series API, we only utilize Lora~\cite{hu2021lora} techniques to conduct instruction tuning for ranking tasks~\cite{bao2023bi} on Llama2 by employing different fairness strategies. The experiments were conducted under four NVIDIA A5000.

For the baseline, we compare four common-used types of methods to mitigate unfairness in LLMs:
(1) \textbf{Self-Align}: following the practice in~\citet{sun2023principledriven}, we utilize ChatGPT-3.5 (stronger LLM) to generate more reliable and fair responses to user's queries and fine-tune the original Llama2 with the high-quality self-aligned responses. (2) \textbf{Re-Weight}: following~\cite{jiang2024item}, during the tuning phase, we set the weight to be inversely proportional to the popularity of the item. (3) \textbf{Data-Augmentation}~\cite{ghanbarzadeh2023gender}: we replace the ``[Demographic]'' placeholder with explicit sensitive attribute as illustrated in Section~\ref{sec:settings}. (4) \textbf{Prompt-Tuning}~\cite{chisca-etal-2024-prompting}: we utilize the prompt-tuning techniques to learn a fairness prompt to decrease the unfairness behaviors.

\subsection{Experimental Results}
In our experimental results, we mainly compare the unfairness of the most common sensitive attributes: gender and race. For the continent, we also observe a similar tendency.

In Table~\ref{tab:main_exp}, it becomes evident that our method significantly outperforms the baselines across all datasets and sensitive attributes, encompassing different top-K ranking sizes. The experiments conclusively demonstrate that our method can mitigate the implicit ranking unfairness effectively.

Meanwhile, in Table~\ref{tab:main_exp}, we observe that our model maintains a high level of effectiveness, with only a small drop in accuracy (around 1-2\%) compared to the best-performing baselines, while achieving significantly better fairness with an improvement of nearly 10-20\% for most cases. These findings indicate that our model strikes an excellent trade-off between fairness and effectiveness, addressing your concerns effectively.

\subsection{Experimental Analysis}
In this section, we will analyze why our method can mitigate implicit ranking unfairness. In Figure~\ref{fig:tsne}, we use TSNE\cite{van2008visualizing} to reduce the dimensionality of the vectors and calculate the distances between them to assess whether the large model reduces the distance between different groups of sensitive attributes in the ranking task.

From Figure~\ref{fig:tsne}, we can observe that using implicit attributes for data augmentation not only reduces the embedding distances between different implicit attributes but also brings embeddings of explicit attributes (such as ``Male, Female'') closer together. In this way, the LLM-based ranking model will find it difficult to infer demographic attributes from user names, thereby effectively achieving ranking fairness.

% Appendix~\ref{app:xxx} shows an illustrative example to show how the interaction of names of users and their demographic features leads to unfairness in the context of ranking.
As for other experiments, Appendix~\ref{app:case_study} shows a case study to show how the interaction of names of users and their demographic features leads to unfairness in the context of ranking.

Meanwhile, to further investigate the intrinsic bias of LLMs, we will conduct experiments to analyze the intrinsic bias of LLMs by removing the history information/names in the ranking prompt to compare the ranking performances with and without browsing history/names. The experiments are shown in Appendix~\ref{app:bias_of_llms}.

\section{Related Work}

Recently, researchers have discovered that LLMs can exhibit discriminatory behaviors~\cite{gallegos2023bias}. In previous discrimination evaluation settings, researchers often measure stereotype sentence pairs that only differ in the sensitive attribute. For example, they often adapt terms ``Male'' and ``Female''~\cite{nangia2020crows, delobelle2022measuring, gallegos2023bias} and for Race, they often substitute terms ``Black'', ``White'' and ``Asian''~\cite{zhang2023chatgpt, tamkin2023evaluating}. 
Among the allocational harms, previous studies found that LLMs often exhibit discrimination against certain groups. For example, ~\citet{salinas2023unequal, de2021stereotype, mcgee2023gender, thakur2023language, bolukbasi2016man} discovered that LLMs will generate discriminatory content for disadvantaged gender. ~\cite{zhang2023chatgpt} show recommendation outcomes may discriminate against certain groups, see also~\cite{rozado2023political, hutchinson2020social}. In our research, we mainly utilize the counterfactual fairness concept to measure the \textit{implicit ranking unfairness} of LLMs-based recommendation.

There are some works that try to mitigate unfairness problems in LLMs. For example, RLHF~\cite{ouyang2022training} and RLAIF~\cite{bai2022constitutional} try to utilize reinforcement learning to align LLMs with human values. Generally, to address the imbalance in the original dataset against certain groups, some work~\cite{ghanbarzadeh2023gender,zhang2023chatgpt, lu2020gender} create matched pairs (\eg male or female) to ensure a more equitable dataset and other methods~\cite{dixon2018measuring, sun2022moraldial} add non-toxic examples for groups. Other approaches~\cite{orgad2022blind, deldjoo2024cfairllm} suggest the use of down-weighting samples containing social group or discriminated information as a re-sampling strategy. While some method proposes to utilize the prompt-tuning method to learn a fair-aware prompt~\cite{hua2023up5, chisca-etal-2024-prompting}. Moreover, other studies~\cite{raffel2020exploring, ngo2021mitigating} propose to filter out and remove discriminated or taxonomic content from datasets.

\section{Conclusion}
%In summary, we uncover Goodhart’s Law in evaluating discrimination of LLMs: prior evaluation metrics will gradually the underestimate discrimination degree when utilizing common demographic terms.
%Then we introduce an IV-based evaluation process for assessing the discrimination and suggest user names as a suitable IV for evaluating demographic discrimination behaviors. Experimental results show that ChatGPT exhibits discrimination against certain groups by nearly 2-4 times more than in previous evaluations, posing a significant threat to the ethical usage of LLMs.

In conclusion, our findings show that LLMs exhibit serious implicit ranking unfairness. This implies that, even when sensitive attributes are not explicitly provided, LLMs can still exhibit discriminatory ranking behaviors. Regarding the root causes, we find that LLMs’ capability to deduce sensitive attributes from non-sensitive attributes contributes to intrinsic discriminatory knowledge. Finally, we propose to mitigate such unfairness effectively by utilizing fair-aware data augmentation. We emphasize the necessity of identifying and moderating implicit ranking unfairness in existing LLMs.

%to avoid potential reinforced human-LLMs ecosystem deterioration.

\section*{Limitations}

%owever, over time, it is possible that user names may also attract attention, for example, the LLMs will also filter the user names in the later version. How to identify an effective instrumental variable (IV) remains a crucial yet unresolved challenge in economics~\cite{baiocchi2014instrumental}. 

% Secondly, in this paper, we verify that confounder creates a backdoor path between commonly used demographic terms and evaluation metrics. The exact nature of the confounder remains unknown. Here, we only make an assumption: previous work shows that RLHF~\cite{ouyang2022training} and RLAIF~\cite{bai2022constitutional} have been effective to alleviate user discrimination, and many works~\cite{karwowski2023goodhart, ashton2020causal} have shown that one potential scenario of Goodhart's Law in evaluating discrimination may involve embedding it in the reinforcement learning framework of RLHF. However, it still needs a detailed analysis to find out the nature of the confounder.

Finally, in our paper, we mainly utilize ChatGPT, and Llama2 as our evaluation 
LLMs and only test the discrimination behaviors against demographic information in recommendation tasks. Meanwhile, we currently only select user names and user emails as the implicit attribute. However, different LLMs and different discrimination behaviors may exhibit different forms of implicit unfairness. This paper serves as a valuable illustration to the community, emphasizing the importance of careful consideration when assessing the discrimination behaviors in LLMs.

\section*{Ethics Statement}
This study is a retrospective analysis conducted on publicly available datasets with research-oriented licenses, involving neither human participants nor the requirement for informed consent. All results generated by LLMs are utilized for offline analysis by the authors and remain invisible to real-world users, ensuring no actual social impact. User profiles used in the experiments, including names, genders, races, and nationalities, are simulated, and all user identities have been completely anonymized. The primary objective of this study is to enhance the fairness of LLMs, aligning with the principles of responsible and ethical usage.

\section*{Acknowledgements}
This work was funded by the National Key R\&D Program of China (2023YFA1008704), the National Natural Science Foundation of China (No. 62377044, 62276248), the Youth Innovation Promotion Association CAS (No. 2023111), Beijing Key Laboratory of Big Data Management and Analysis Methods, Major Innovation \& Planning Interdisciplinary Platform for the  ``Double-First Class” Initiative, Public Computing Cloud, funds for building world-class universities (disciplines) of Renmin University of China. Supported by the Outstanding Innovative Talents Cultivation Funded Programs 2024 of Renmin University of China.

\bibliography{custom}
\clearpage
\appendix

\section*{Appendix}
\section{Symbol Descriptions}
\label{sec:symbol}
Given that some symbol definitions may be complex, we provide the notations in Table~\ref{tab:symbol} to help readers better understand them.

\begin{table}[t]
\small
\caption{Detailed symbol definitions. }
\label{tab:symbol}
\centering
\begin{tabular}{|ll|}
symbol & explain \\
\hline
$u$    & user    \\
$i$    & item    \\
$v_u$    & user $u$'s non-sensitive attribute (e.g. names)    \\
$s_u\in \mathcal{S}$    & user $u$'s sensitive attribute (e.g. male)    \\
$p_u$    & personalized prompt for each user $u$    \\
$K$    & ranking size    \\
$L_K(u)$    & ranking list to user    \\
$\mathcal{N}_s, s\in \mathcal{S}$    & user names set belong to attribute $s$    \\
$e(\cdot)$    & function to map text to embedding    \\
$T_i$    & i-th topic sentence    \\
$S_i$    & similarity score of i-th topic    \\
$\hat{z}$    & predicted attribute similarity distribution    \\
$(i_j , i_n)$    & item pair from ranking list    \\
\end{tabular}
\end{table}

\section{Topic Sentences}\label{app:topic_sentences}

The topic sentences $T_i$ consist of the keywords of certain topics. 
In the recommendation task, we choose six topics for comparison using the following prompt: \textit{Please give me 20 keywords related to the {{topic}} news/jobs}.

After filtering some neural words, the news keywords associated with the topics are listed below.
\begin{itemize}
    \item Politic: violence, elections, government, legislation, political, diplomacy,  corruption, democracy, voting, legislation, trump, hurricane
    \item Life: dog, technology, travel, food, finance, environment, weather, transportation, relationships, family, career, hobbies, events, shopping
    \item Education: school, university, teacher, student, curriculum, exams, educational, scholarships, literacy, academic
    \item Health: health, fitness, pandemic, vaccine, medication, fat, sleep, nutrition, exercise, diet, death
    \item Art: art, venice, gallery, artist, exhibition, painting, sculpture, museum, culture, fashion, entertainment, auction, design
    \item Sports: sports, football, game, team, coach, basketball, baseball, swimming, athletics, exercise
\end{itemize}

After filtering some neural words, the job keywords are listed as follows.

\begin{itemize}
    \item Service: Customer, Service, Sales, Associate, Receptionist, Waiter, Waitress, Hotel, Concierge, Flight, Attendant, Cook, Housekeeper, Lifeguard
    \item Health$\&$Medical: Health, Medical, doctors, nurses, surgeons, medical, technicians, pharmacists, healthcare
    \item Business$\&$Finance: Business, Finance, management, finance, accounting, marketing, entrepreneurship, administrater
    \item Education$\&$Teaching: Education, Teaching, teachers, professors, tutors, school, librarians, educational, counselors
    \item Engineering$\&$Technical: information, technology, computer, science, programming, software, development, network, administration, data, analysis, electrical
    \item Arts$\&$Entertainment: arts, media, entertainment, actors, musicians, writers, filmmakers, designers, photographers, artists
\end{itemize}

\section{Case Study}\label{app:case_study}
In this section, we give a case study how user names interact with the sensitive attribute. We test three items (news) as the ranking [Candidate]:
\begin{itemize}
    \item A: Some believe Mason Rudolph, hit in the head with his own helmet, isn't getting enough blame. 
    \item B: Taylor Swift Rep Hits Back at Big Machine, Claims She's Actually Owed 7.9 Million in Unpaid Royalties. 
    \item C: This is it, this is the luckiest break in the history of golf.
\end{itemize}

Then, we test gender discrimination by utilizing the male name \textit{Jack} and the female name \textit{Sophie}. The ranking results are ``A,B,C'' nad ``B,A,C'', respectively.

By analyzing the word embedding by Llama2, we find that the embedding of word \textit{Jack} is more close to Male: the word \textit{Jack} has 56\% similarity with \textit{Male} and 44\% with \textit{Female}). While \textit{Sophie} is more close to \textit{Female}: about 54\% similarity with \textit{Male} and 46\% with Male. Therefore, from the ranking result, we can observe user named Jack is more likely to rank male-related news (A) in a higher position, and the user named Sophie is more likely to rank female-related news (B) in a higher position. Such intrinsic bias embedded in LLMs leads to the implicit ranking unfairness behaviors of LLMs.

\section{Intrinsic Bias of LLMs}\label{app:bias_of_llms}
In this section, to further investigate the bias is only from the intrinsic bias of LLMs but not from browsing history, we will conduct experiments to analyze the intrinsic bias of LLMs by removing the history information/names in the ranking prompt to compare the ranking performances with and without browsing history/names. 

\begin{figure}[t]
    \centering
     \includegraphics[width=\linewidth]{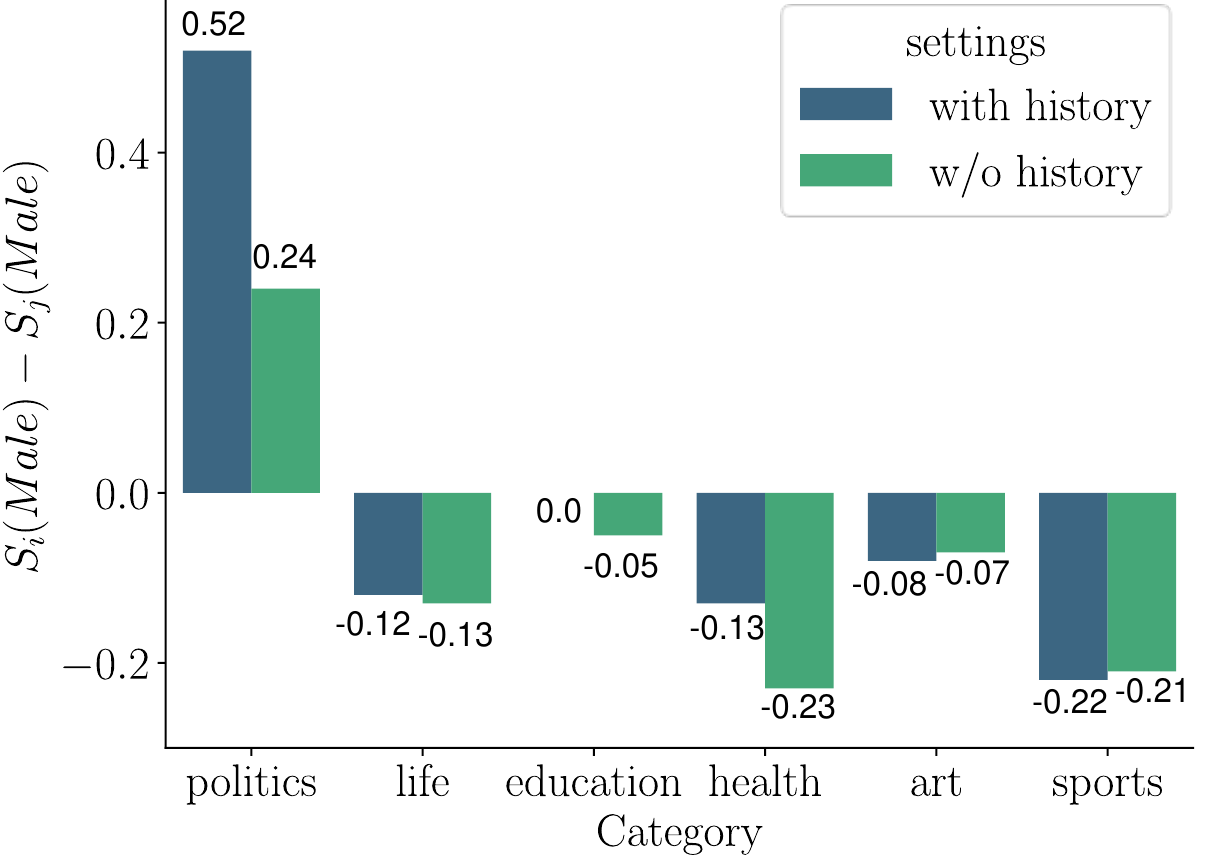}
    \caption{Topic similarities comparison with and without using the browsing history.}
    \label{fig:removing_history}
\end{figure}

\subsection{Removing History}
Firstly, we conducted the experiment only based on user names by utilizing the following prompt: "Please give five news/jobs ranking list to the user named [names]". Then, we analyze the topic distribution with and without using the browsing history to investigate whether their discrimination patterns are similar, assessing the extent of the browsing history's influence. The experiments were conducted under the MIND dataset and gender-based sensitive attributes.

Following the settings in Section~\ref{sec:settings}, we compare the topic similarity gap between Male and Female: $S_i(Male)-S_j(Male)$, where $S_i(\text{gender})$ is the $i-$th topic similarity under gender names. We report the distribution value on genders in Figure~\ref{fig:removing_history}. From the data in Figure~\ref{fig:removing_history}, the two data groups have a Pearson correlation coefficient of 0.966, indicating a significant positive correlation.  The experiments demonstrate that even when we remove the history and only utilize user names, the models still exhibit similar patterns of discriminatory ranking, confirming the intrinsic bias present in LLMs.

\subsection{Removing User Names}
In this section, we aim to remove the user names in the prompts to observe the bias from the user's historical browsing history. We compare the ranking performances (NDCG@3) with and without user names in Table~\ref{tab:removing_usernames}. The experiments were conducted under both the news and job dataset and gender-based sensitive attributes.

\begin{table}[t]
\caption{Ranking accuracy for removing user names. The ranking metric is NDCG@3 and ``w/o names'' denotes removing user names only contains user historical behaviors. }
\label{tab:removing_usernames}
\small
\centering
\begin{tabular}{cccc}
\hline
domain & w/o names & male names   & female names \\
\hline
news    & 0.460  & 0.463 & 0.659  \\ 
jobs    & 0.495  & 0.500 & 0.502  \\ 
\hline
\end{tabular}
\end{table}

From the table, we can observe that compared to the situation without names as input, the inclusion of users' names significantly amplifies the disparities in treatment among different groups, thereby verifying the intrinsic bias present in LLMs.

\end{document}